\documentclass[aps,pre,twocolumn,groupedaddress,floatfix,superscriptaddress]{revtex4}
\usepackage{dcolumn}
\usepackage{amsmath}
\usepackage{amssymb}
\usepackage{lipsum}
\usepackage{graphicx}
\usepackage{bm}
\usepackage[T1]{fontenc}
\usepackage{color}
\usepackage{url}
\usepackage[bf]{subfigure}
\usepackage{rotating}
\usepackage{soul}
\usepackage{scalefnt}
\usepackage{hyperref}
\usepackage{scalefnt}

\usepackage{perpage} 
\MakePerPage{footnote}

\usepackage{color}
\definecolor{redcolor}{rgb}{1.0,0.,0.}

\usepackage[utf8]{inputenc}

\begin{document}
\title{Intelligent Complex Networks}

\author{Henrique F. de Arruda}
\affiliation{Institute of Mathematics and Computer Science, University of S\~ao Paulo, S\~ao Carlos, SP, Brazil.}
\author{Cesar H. Comin}
\affiliation{Department of Computer Science, Federal University of S\~ao Carlos - S\~ao Carlos, SP, Brazil}
\author{Luciano da F. Costa}
\email{ldfcosta@gmail.com}
\affiliation{S\~ao Carlos Institute of Physics, University of S\~ao Paulo, S\~ao Carlos, SP, Brazil}

\begin{abstract}
The present work addresses the issue of using
complex networks as artificial intelligence
mechanisms.  More specifically, we consider the
situation in which puzzles, represented as
complex networks of varied types, are to be 
assembled by complex network processing 
engines of diverse structures.  The puzzle
pieces are initially distributed on a set of nodes chosen according to different criteria, including degree and eigenvector centrality.  The pieces are then repeatedly copied to the neighboring nodes.  The provision of buffering of different sizes are also investigated. Several interesting results are identified, including the fact that BA-based assembling engines tend to provide the fastest solutions. It is also found that the distribution of pieces according to the eigenvector centrality almost invariably leads to the best performance.  Another result is that using the buffer sizes proportional to the degree of the respective nodes tend to improve the performance.
\end{abstract}
\maketitle

\setcounter{secnumdepth}{1}

\section{Introduction}

Human beings enjoy, and often excel at solving puzzles.  Endowing machines with comparable reasoning abilities has provided motivation for generations of scientists.  Yet, while machines have overtaken us in several abilities, such as calculating and remembering, automatic reasoning remains a big challenge~\cite{davis2016scope}.  The considerable body of related studies is part of the broad area known as \emph{Artificial Intelligence - IA}.  Indeed, a vast number of different principles and approaches have by now been tried, ranging from formal grammars (e.g.~\cite{al2017adaptive,fang2018learning}) to artificial neuronal networks (e.g.~\cite{schmidhuber2015deep,sainath2015deep}).  The recent advent of \emph{network science}~\cite{barabasi2016network} has contributed valuable new concepts and methods that can be tried in IA. Some related works involve neuronal models and simulations~\cite{de2015framework}, studies of knowledge acquisition~\cite{de2017knowledge}, jigsaw percolation~\cite{brummitt2015jigsaw}, applications to neural networks~\cite{stauffer2003efficient}, and other machine learning approaches~\cite{stam2014modern}.

One of the remarkable features of complex networks is their ability to represent, model and simulate virtually any discrete system and phenomenon, including artificial intelligence.  In addition, network models exist that are able to incorporate several remarkable topological features such as power law degree distribution~\cite{barabasi1999emergence}, small world interconnectivity~\cite{watts1998collective}, and assortativeness~\cite{newman2002assortative}.  Moreover, the knowledge about the effect of these varying topologies on the unfolding of dynamics in networks, with dynamical states associated to nodes, has been growing steadily~\cite{da2007correlations,comin2013relationship,souza2018topology}. All in all, network science provides new insights and methods for all the main aspects involved in AI research, namely: (1) representing a problem/solution; (2) representing the topology of reasoning machines; (3) providing ways to combine and integrate parts of the solution; and (4) addressing different interactions between the topology and dynamics of problem solving. 

\begin{figure}[!ht]
 \centering
  \includegraphics[width=.45\textwidth]{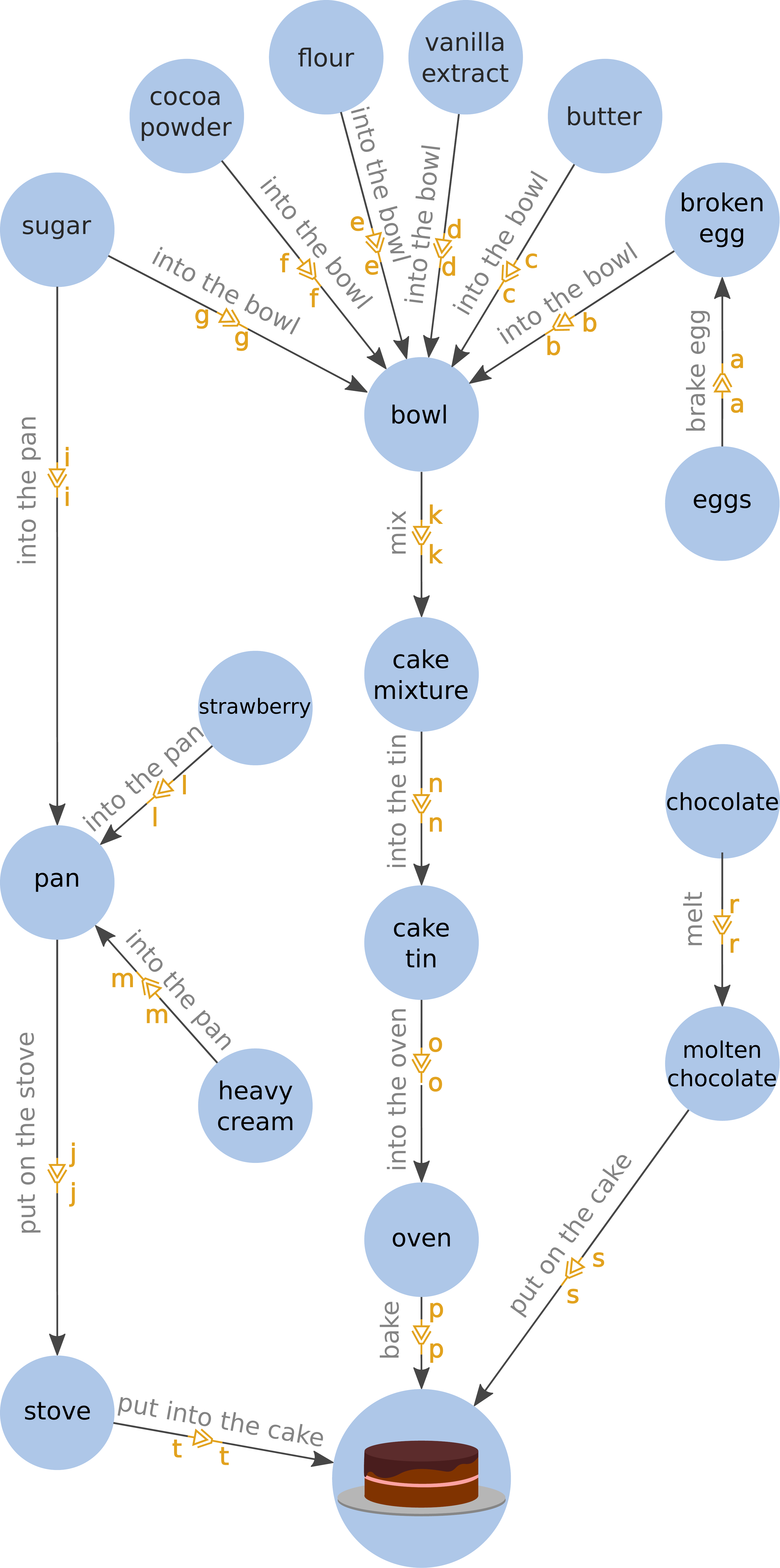}
 \caption{Toy example of a body of knowledge respective to a cake recipe being represented as a graph $G$. The network nodes represent the necessary ingredients and kitchen utensils, and the edges indicate the actions involved in making a cake.  Also shown are  tags (letters shown in ochre) that can be associated to each of the edges of $G$ in order to uniquely identify each of the links.}
 \label{fig:cake}
\end{figure}

The present work aims at applying as much as possible the above outlined resources from network science in order to derive a potentially new general model of reasoning machines.  More specifically, we consider that several bodies of knowledge in the real world can be effectively represented in terms of complex networks whose nodes correspond to concepts while the links stand for relationships between these concepts.  Figure~\ref{fig:cake} provides a toy example of this type of representation.  For generality's sake, we consider that the body of knowledge can follow several theoretical complex network models, including BA, WS and ER.

Given a body of knowledge represented as a complex network, we proceed to investigate the problem of how reasoning engines underlain by complex networks of several types (i.e.~BA, WS and ER) perform while assembling the given body of knowledge.  More specifically, the sought solution of a problem is represented as a complex network composed of basic concepts (nodes) that are interrelated by specific edges, each labeled with respective tags.    This solution is then decomposed into isolated nodes, but retaining the tagged connecting spokes.  These basic components of the solution are henceforth called \emph{nodelets}.   Figure~\ref{fig:scheme2} illustrates a simple solution and its respective decomposition into nodelets.  Another network, the reasoning engine (\emph{RE}), is responsible for assembling the puzzle.  The nodelets derived from the solution are progressively distributed among the nodes of the RE (also illustrated in Fig. 1), as the solution of the problem requires the junction of all nodelets while respecting their interconnectivity tags.  

So, the task of problem solution is posed here in terms of a network (the reasoning engine) attempting to recover a network (the solution).  Several dynamics can be considered for obtaining the solution, involving different ways to distribute the nodelets, communication between nodes in the RE, assigning merits to each putative solution, incorporating forgetting mechanisms, and heuristics for seeking for missing parts.  In this work, we adopt what we  believe is one of the simplest solution approaches: (i) each node keeps a list of so far received nodelets. (ii) at each time step, each node receives the lists from all its neighbors and merge it into its own list, removing replications; (iii) matching parts in the current lists of each node are linked, and the list updated.   In this way, the solution is progressively assembled at each of the nodes.   After a number of steps corresponding to the diameter of the RE, all its nodes will have assembled the solution.  This framework paves the way for a large number of investigations, in the present work we focus at the progress of the assembling with respect to the original solution and RE following several complex networks models (Erdos-Rényi, Barabási-Albert, and Watts-Strogatz).

To complement the study, we also consider different strategies for initially distributing the nodelets among the RE nodes. We use two centrality measurements, the node degree, which is a local measurement, and the eigenvector centrality, which is computed globally for the network. These two measurements were considered in increasing and decreasing order. We also distributed the nodelets by random choice.

Though intrinsically simple, the above outlined approach reflects a large number of features found in problem solution by a group of people, the nervous system, and even the unfolding of science.  Indeed, the solution of a problem almost always involve the combination of several parts into an integrated whole.  


\section{Review of Used Concepts and Methods}

Here, we considered three types of network models. The first one is the model called Erd\H{o}s-R\'enyi (ER)~\cite{erdos1960evolution}. An ER network starts with a given number of nodes, and the connections are created randomly, according to a probability $p$.  Another model, incorporating the characteristics of the power-law distribution of node degrees, is based on preferential attachment, being known as the Barab\'asi-Albert model (BA)~\cite{barabasi1999emergence}.  More specifically, the network is created by adding nodes that connects to $m$ other nodes with probability proportional to the node degree. We also considered the Watts-Strogatz (WS) approach, in which the networks are created from a toroidal lattice and the edges are subsequently rewired according to a probability $p$~\cite{watts1998collective}. Note that in the original version of WS model, the initial structure is based on nodes distributed along a ring, but here we consider a 2D lattice. Furthermore, each node is initially connected to the four nearest neighbors.

In addition to these network models, it is also interesting to adopt networks considering spatial information, i.e.~ geographical network models. A possible choice is the Waxman model (WAX)~\cite{waxman1988routing}, which consists in connecting the nodes according to a probability that exponentially decreases according to the geographical distance between nodes, that is:
\begin{equation}
p = \alpha \exp{(-d/\beta)},
\end{equation}
where $d$ is the distance and $\alpha$, and $\beta$ are constants. In order to create networks with a fixed average node degree, we randomly draw the WAX edges until the desired average degree is reached.

\section{Experimental framework}
\begin{figure*}[!ht]
 \centering
  \includegraphics[width=1.\textwidth]{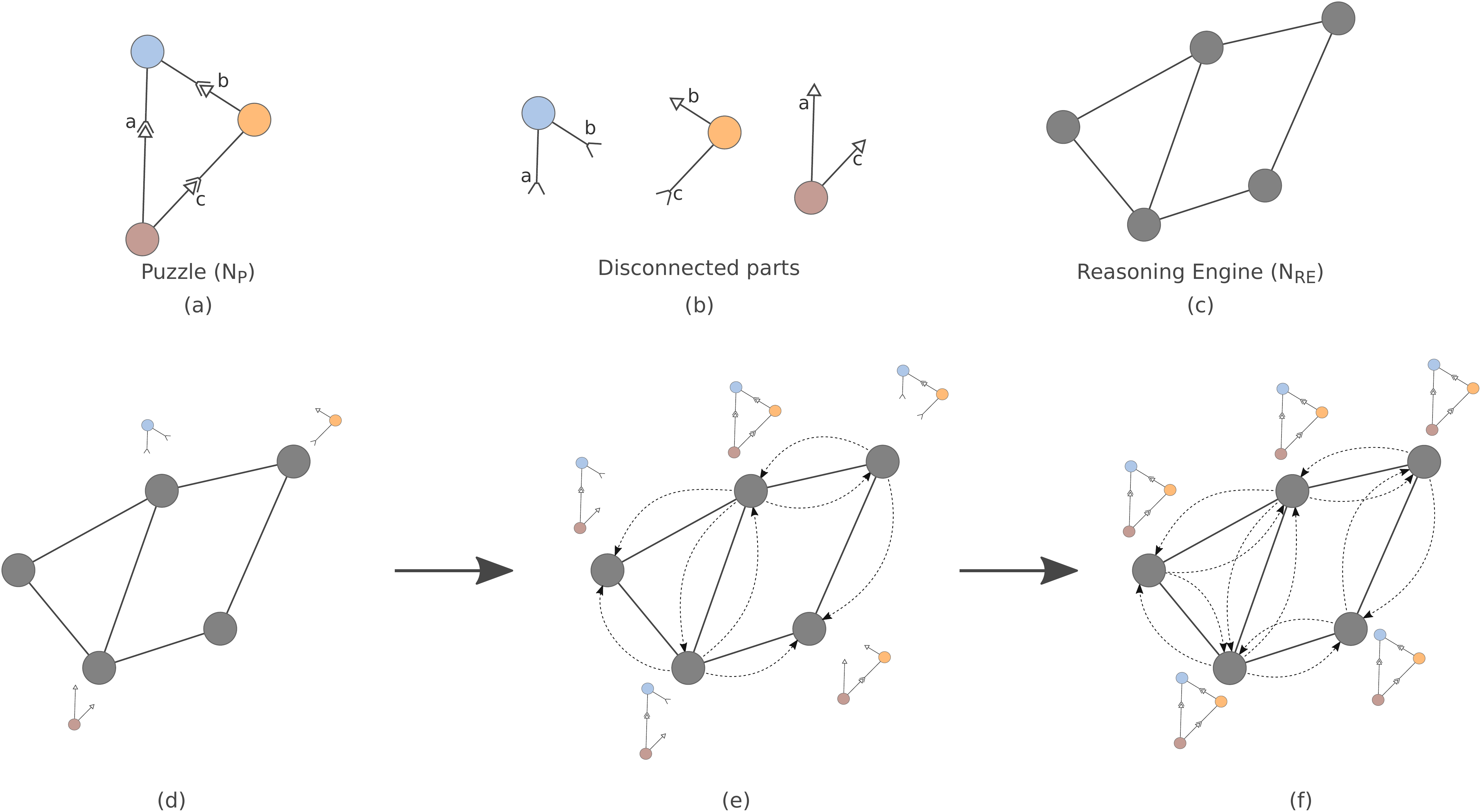}
 \caption{Example of the one execution of the dynamics. The first row shows the necessary elements to execute the dynamics and the second row shows an example of execution.}
 \label{fig:scheme2}
\end{figure*}

As illustrated in Figure~\ref{fig:scheme2}, the dynamics studied in this work consists of assembling a puzzle in all nodes of a network. Two  networks have been considered: one representing the puzzle ($N_P$) Figure~\ref{fig:scheme2}(a), and another, the RE network ($N_{RE}$) Figure~\ref{fig:scheme2}(c), corresponding to the reasoning engine responsible for solving the puzzle. In order to assemble the puzzle, each of the nodes in $N_{RE}$ tries to assemble all possible $N_P$ nodelets.  The dynamics starts with a set of disconnected nodelets Figure~\ref{fig:scheme2}(b). The nodelets have specific tags, which shall be connected by a $N_{RE}$ node to the respective tag from another nodelet. The dynamics starts by assigning each nodelet to a respective node in the $N_{RE}$ network (Figure~\ref{fig:scheme2}(d)).  This is done according to a given complex network measurement (e.g. the degree) computed on the $N_{RE}$ network. For instance, the nodelets can be assigned to the $N_{RE}$ nodes with the highest degrees. This procedure is henceforth called the \emph{assignment priority} (AP).

For each iteration of the dynamics, the $N_{RE}$ nodes send a replica of their nodelets to all their respective neighbors, which also store and join all the possible parts (Figure~\ref{fig:scheme2}(e)). 

In addition to considering the assigned priority, we introduce a limitation in $N_{RE}$, which consists in defining a buffer that can store a fixed number of $N_P$ components. In other words, when $N_{RE}$ have more $N_P$ components than the buffer size, only the largest components are kept. Cases where components with the same size are to be maintained or eliminated are solved by randomly selecting the components for removal. The experiments were divided into two approaches. The first considers the same buffer size for all nodes of $N_{RE}$, and in the second, the buffer size was defined as the node degree of $N_{RE}$.

\subsection{Random Nodelet Distribution}
There are many possibilities to choose the assignment priority of the of $N_P$ nodelets in $N_{RE}$, and one of the most straightforward manner is to apply a random selection. Note that in the assignment priority, each of the nodes of $N_{RE}$ can receive a single $N_P$ nodelet. 

\subsection{Degree-Based Nodelet Distribution}
In order to better understand how the assignment priority modify the conditions to mount the puzzle ($N_P$), we considered centrality measurements of networks. Similarly to the previous case, each node of $N_{RE}$ can receive a single $N_P$ nodelet, but here the assignments are chosen according to a given centrality measurement. The first of the considered centrality measurements was the node degree ($k$). We considered this measurement in increasing and decreasing order. 

\subsection{Eigenvector-Based Nodelet Distribution}
In the previous case, we considered the degree centrality, which reflects only the local information of a node. In order to incorporate global information about the networks, in terms of all nodes, we employed the eigenvector centrality ($EC$)~\cite{bonacich1987power}. This centrality measurement consists in iteratively giving to each node a score that is proportional to the sum of the score of its neighbors~\cite{newman2010networks}. The eigenvector centrality can be computed in terms of the eigenvector associated with the largest eigenvalue of the adjacency matrix. Similarly to $k$, the $EC$ values of $N_{RE}$ were considered in increasing and decreasing order. 

\section{Discussion}
In this section, we present the results and discuss them. This discussion is divided into two cases. In the first case, we show the findings regarding the original framework, in which there is no limitation for the storage of the $N_{RE}$ nodes. This case is henceforth referred as ``infinite buffer size''. The other case concerns applying the dynamics with limited buffer size, which is henceforth called ``limited buffer size''.

\subsection{Case 1: Infinite Buffer Size}
In the following, we present the results obtained for the several assignment priorities, and discuss the differences between them. One characteristics that can vary substantially among the considered networks is the diameter, which is defined as the largest shortest path between each pair of nodes. Note that the network diameter limits the maximum number of iterations of the dynamics. As a consequence of this characteristic, there is a tendency for networks with a lower diameter to finish all the puzzles sooner. To be able to compare among $N_{RE}$ networks with different diameters, we divided the time by the network diameter, henceforth called normalized time. 

\begin{figure*}[!htpb]
  \centering
    \subfigure[$N_P =$ WS and $N_{RE} = $ WS]{\includegraphics[width=0.3\textwidth]{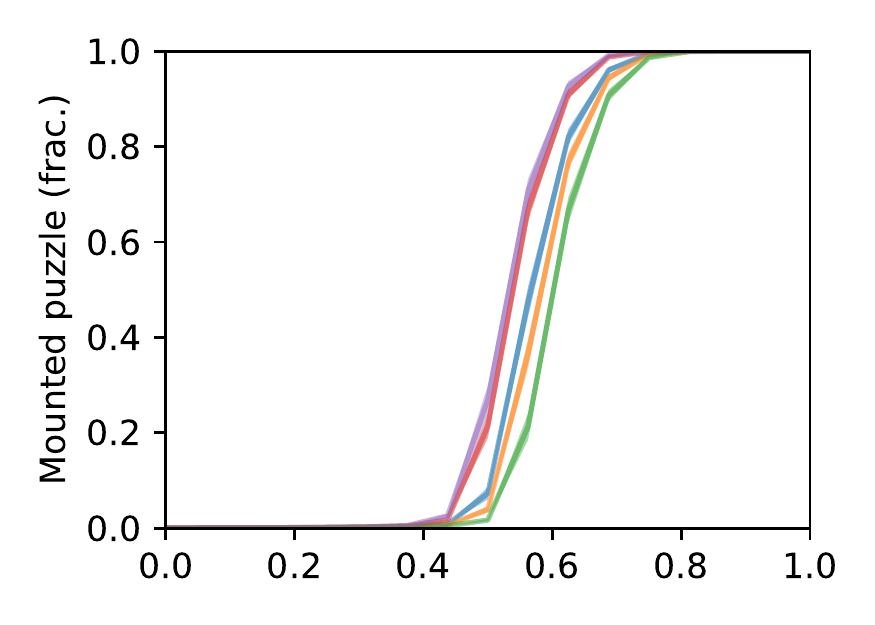}}
    \subfigure[$N_P =$WS and $N_{RE} = $ ER]{\includegraphics[width=0.3\textwidth]{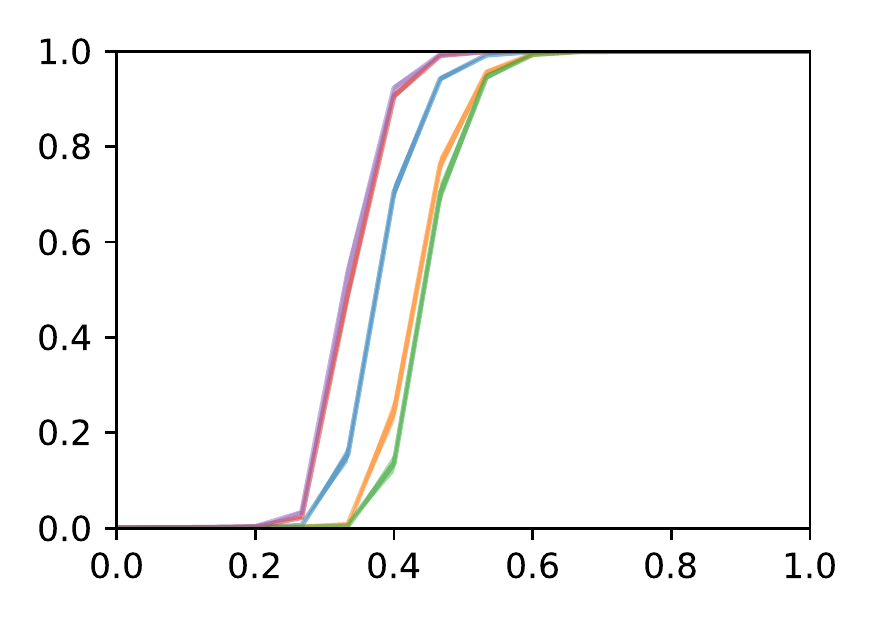}}
    \subfigure[$N_P =$WS and $N_{RE} = $ BA]{\includegraphics[width=0.3\textwidth]{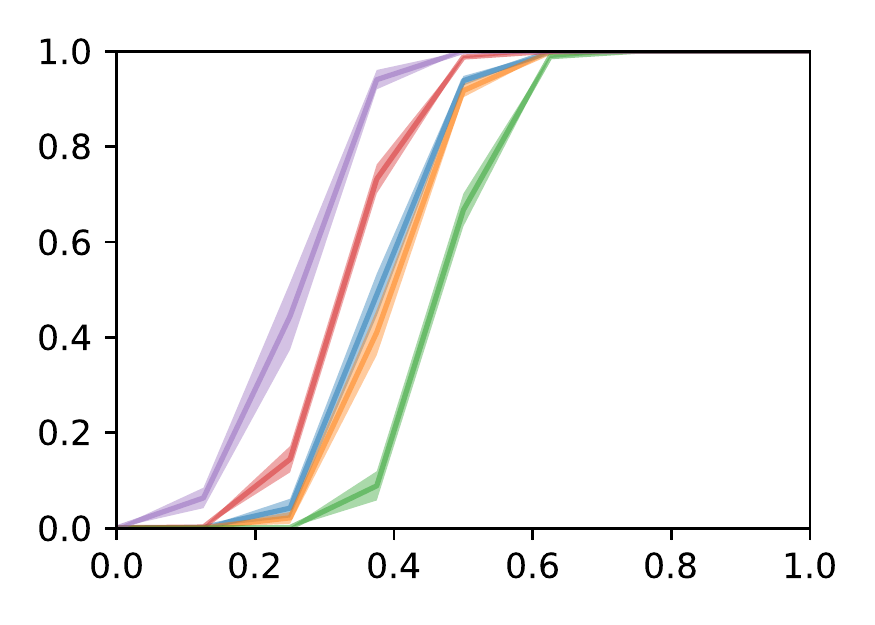}}
    \subfigure[$N_P =$ER and $N_{RE} = $ WS]{\includegraphics[width=0.3\textwidth]{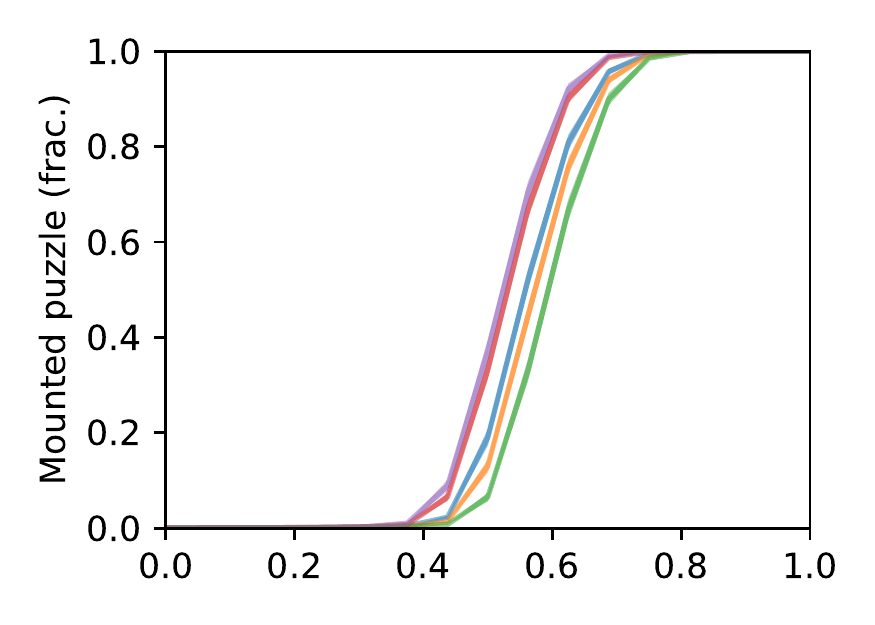}}
    \subfigure[$N_P =$ ER and $N_{RE} = $ ER]{\includegraphics[width=0.3\textwidth]{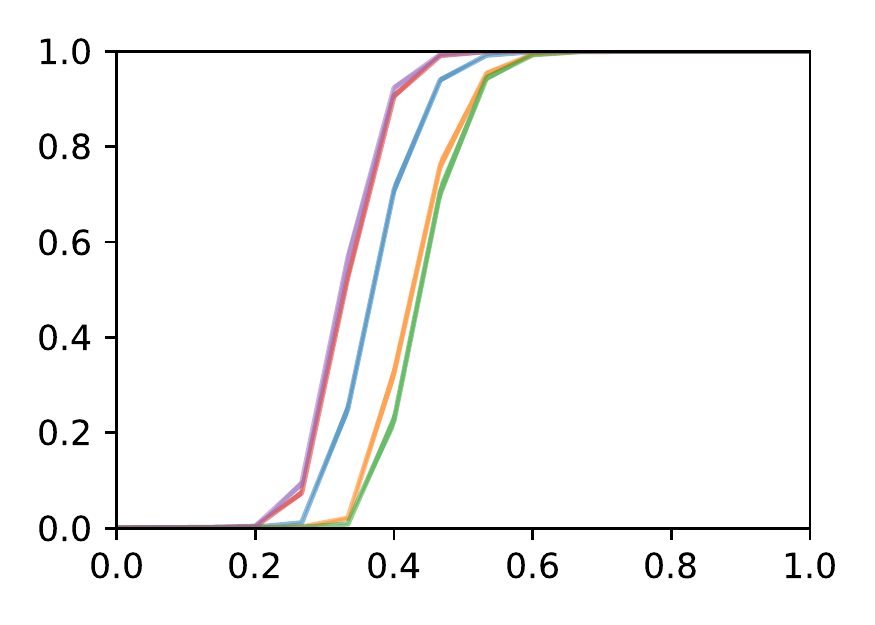}}
    \subfigure[$N_P =$ ER and $N_{RE} = $ BA]{\includegraphics[width=0.3\textwidth]{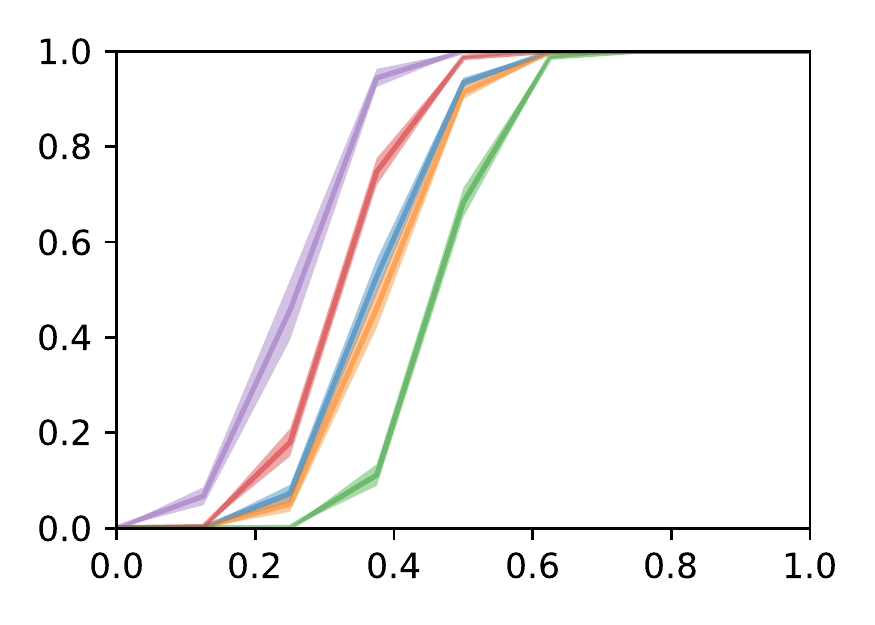}}
    \subfigure[$N_P =$ BA and $N_{RE} = $ WS]{\includegraphics[width=0.3\textwidth]{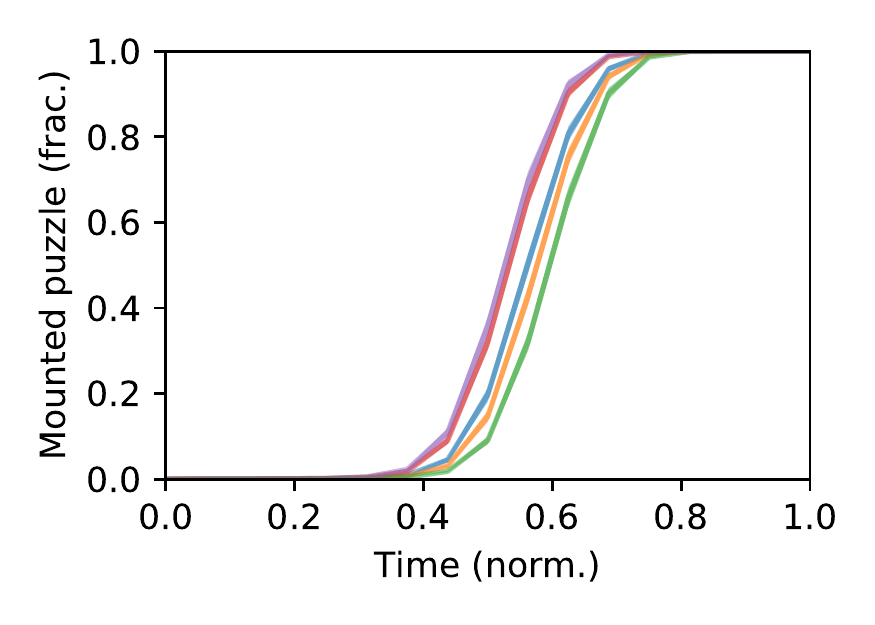}}
    \subfigure[$N_P =$ BA and $N_{RE} = $ ER]{\includegraphics[width=0.3\textwidth]{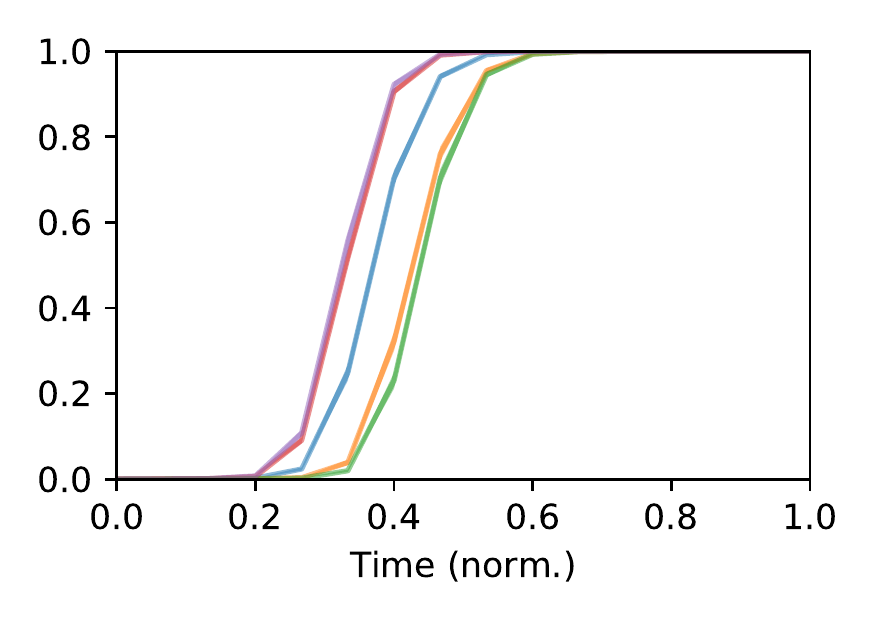}}
    \subfigure[$N_P =$ BA and $N_{RE} = $ BA]{\includegraphics[width=0.3\textwidth]{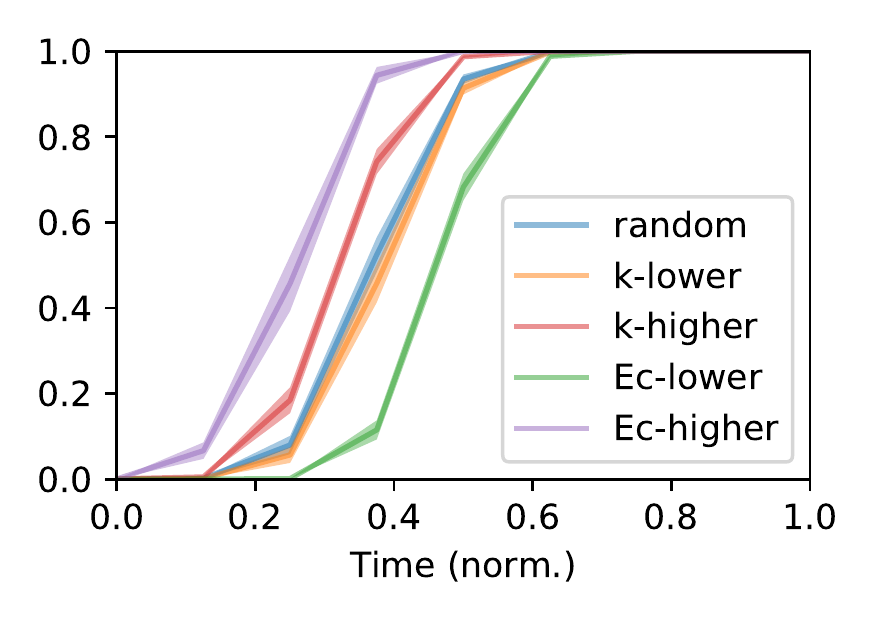}}
  \caption{Each plot shows the fraction of assembled $N_P$ nodes as a function of time. The time axis is normalized by the total amount of time required for all nodes to assemble the puzzle, which is equal to the diameter of the network.}
  \label{fig:mounted}
\end{figure*}

Figure~\ref{fig:mounted} shows the fraction of mounted $N_P$ nodes (calculated as the size of the largest connected component in $N_P$) against the normalized time for all the considered pairs of $N_{RE}$ and $N_P$ networks. The $N_{RE}$ and $N_P$ networks have approximately 10000 and 1000 nodes, respectively. For each test, the simulations were executed 200 times, and for each execution, a new version of each network was generated with fixed parameters, as shown in Table~\ref{tab:parameters}. By comparing the different types of $N_{RE}$ networks, the BA network is the fastest to assemble a significant part of the puzzle, followed by the WS and ER networks. Interestingly, the variation of $N_P$ affected little the fraction of mounted puzzle along time.

\begin{table}[!ht]
\centering
\begin{tabular}{|l|l|l|}
\hline
Network model &$N_{RE}$  & $N_P$      \\ \hline
WS            &$p=0.05  $& $p=0.02$   \\ 
ER            &$p=0.0004$& $p=0.004$  \\ 
BA            &$m=2$     & $m=2$      \\ \hline
\end{tabular}
\caption{Parameters used in the $N_P$ and $N_{RE}$ networks.}
\label{tab:parameters}
\end{table}

Another important characteristic that can be observed in Figure~\ref{fig:mounted} is the assignment priorities of the $N_P$ nodelets. As can be seen, for all considered tests, the assignment priority that sooner provided a significant fraction of mounted puzzles, in increasing order, is given by: higher values of $EC$, higher values of $k$, random nodes, lower values of $k$, and lower values of $EC$. Furthermore, in the majority of the tests, the measured fraction of mounted puzzles was similar between the higher values of $EC$ and $k$ and between the lower values of $EC$ and $k$, in which this difference is slightly more evident when the $N_{RE}$ network is WS. However, a different behavior occurred when BA was employed as $N_{RE}$ because almost all curves were more separated. When the assignment priority was chosen in terms of the lower values of $k$, or at a random order, the curves resulted more similar one another.

\begin{figure*}[!htpb]
  \centering
    \includegraphics[width=1.\textwidth]{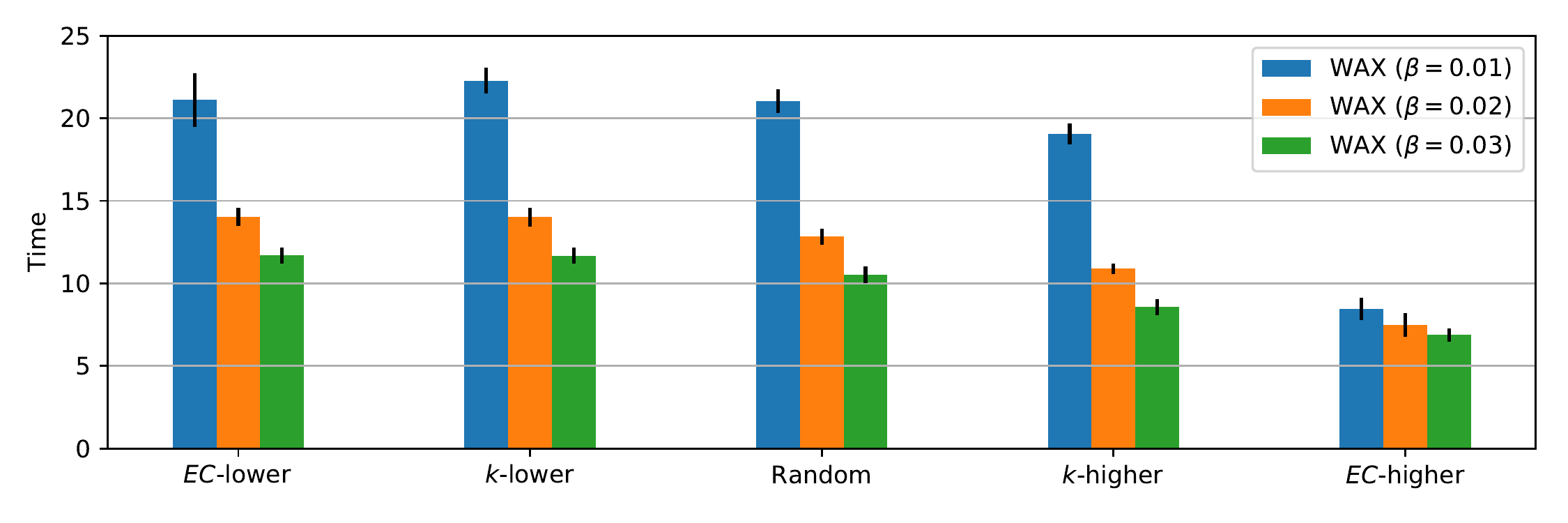}
  \caption{Average time to mount the puzzles ($N_P =$ ER) by considering the variations of WAX networks.}
  \label{fig:wax}
\end{figure*}

As a complementary result, we used the WAX model to obtain networks with varying diameter, and consequently control the shortest path lengths in the networks. The adopted configuration considered: average node degree fixed as 4, approximately 10000 nodes, and $\alpha=1$. For $\beta$ we considered three possibilities: $0.01$, $0.02$, and $0.03$. Note that the shortest path lengths decrease when $\beta$ increases. Figure~\ref{fig:wax} shows $T_F$ (time to mount the first puzzle) for each considered $\beta$. We show only the results for $N_P$ as an ER network because $T_F$ depends only on the number of nodelets. 

The smallest $T_F$ is obtained when the assignment priority is proportional to $EC$. Additionally, by comparing the $N_{RE}$ networks with different values of $\beta$, we found that as $\beta$ increases, the time to mount the puzzles becomes lower. Interestingly, the differences of time to mount the puzzles were smaller when higher values of $EC$ were used to define the assignment priority.

\subsection{Case 2: Limited Buffer Size}

In the following, we considered many different buffer sizes, from 1 to 10, in which the size is the same for all $N_{RE}$ nodes. Because of the computational cost of this variation, here we employed $N_{RE}$ and $N_P$ networks having approximately 1000 and 289 nodes, respectively. For each execution, a new version of each network was generated. The parameters used to generate the networks are shown in Table~\ref{tab:parameters2}. Note that the same assigned priorities employed in the previous case are also considered here, and the dynamics were executed 200 times. 

\begin{table}[!htpb]
\centering
\begin{tabular}{|l|l|l|}
\hline
Network model &$N_{RE}$  & $N_P$      \\ \hline
WS            &$p=0.45$              & $p=0.1$  \\ 
ER            &$p=0.004$             &  $p=0.0135$   \\ 
BA            &$m=2$                     &  $m=2$    \\ \hline
\end{tabular}
\caption{Parameters used for generating the $N_P$ and $N_{RE}$ networks employed on Case 2.}
\label{tab:parameters2}
\end{table} 

\begin{figure}[!htpb]
  \centering
    \includegraphics[width=0.45\textwidth]{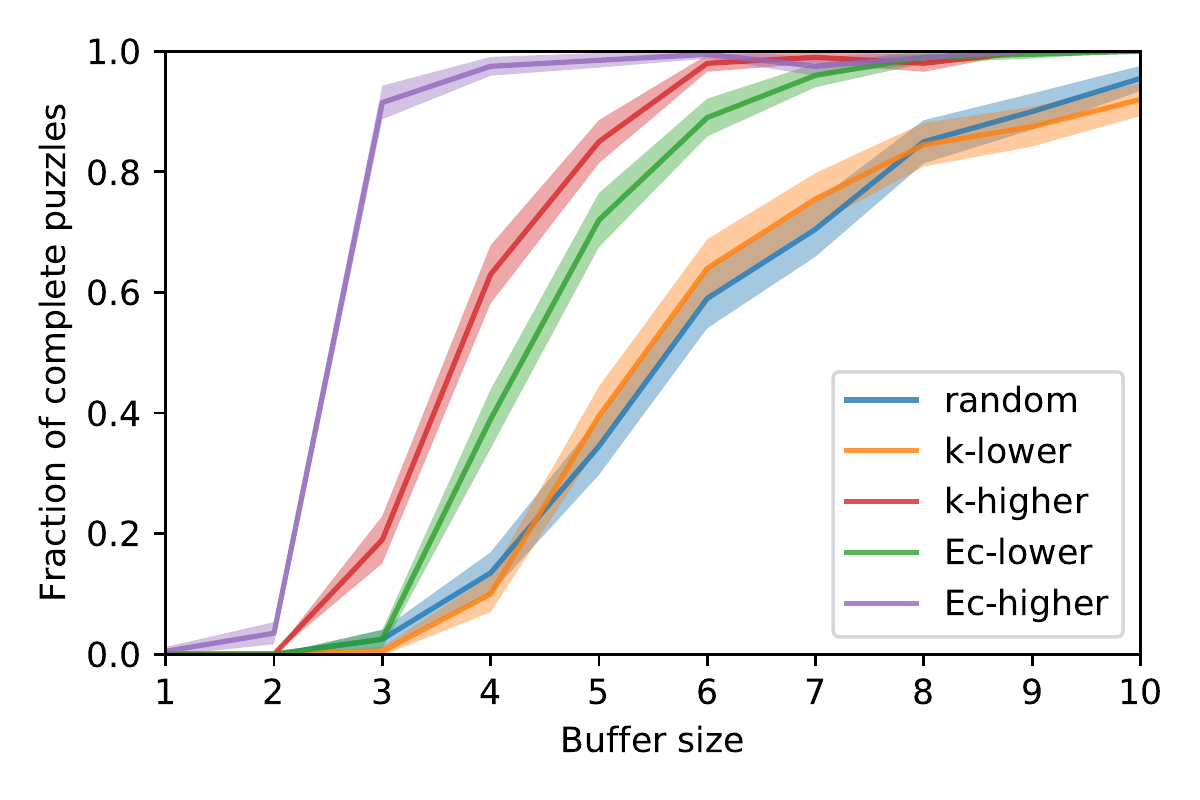}
  \caption{The fraction of dynamics execution that is able to mount the entire puzzle, according to the buffer size. The shaded regions of the curves represent 10\% of the standard deviations. In this example, we use $N_{RE} =$ BA  and $N_P =$ BA. Note that for some sizes of the buffer, the assigned priority of $k$-lower is more effective than the random choice.}
  \label{fig:buffer}
\end{figure}

Figure~\ref{fig:buffer} shows, in terms of the buffer size, the fraction of dynamics executions for which $N_{RE}$ was able to assemble a complete puzzle.  BA $N_{RE}$ and $N_P$ networks were used, but the results are qualitatively similar for the other cases because, for all tests, as the buffer size increases, the fraction $N_{RE}$ of completely mounted cases also increases. However, some differences are found according to the choice of $N_{RE}$. For instance, when $N_{RE} =$ BA, the use of smaller buffer sizes allows a higher number of solutions, for all choices of $N_P$. In the case of the other choices of $N_{RE}$, the employed $N_P$ influences the number of complete solutions. When BA is used as $N_P$, this number is higher for lower values of buffer size. The assigned priorities also depended on the choice of $N_{RE}$. In case $N_{RE} =$BA, there is a clear difference between the curves, except when comparing the $k$-lower and random strategies. For the other cases of $N_{RE}$, the dynamics performed for the assigned priorities of $EC$-lower, $k$-lower, and random also provided similar results.

By considering buffer sizes of 2 and 3, the reasoning engine is rarely able to mount the puzzle. For buffer sizes of three positions, when assigned priority was set as $Ec$-higher, and $N_{RE}$ was BA, the networks were mounted in more than $84\%$ of the cases. Interestingly, by using assigned priority as $k$-higher and $N_P$ as BA, the percentages of entirely mounted $N_P$ are 43\%, 52\%, and 19\% for $N_{RE}$ as WS, ER, and BA, respectively. 

Table~\ref{tab:buffer4} shows the fraction of puzzles mounted when the buffer has a size of 4 components. For all combinations of $N_{RE}$ and $N_P$, the better results are found for $k$-higher and $Ec$-higher. The same pattern is found for the rest of the considered buffer sizes. Furthermore, the better $N_{RE}$ is BA, and the $N_P$ that is most easily mounted is also the BA network.
 
\begin{table*}[!htpb]
\begin{tabular}{l|c|c|c|c|c|c|c|c|c|c|c|c|c|l|l|}
\cline{2-16}
 & \multicolumn{3}{c|}{Random} & \multicolumn{3}{c|}{$k$-lower} & \multicolumn{3}{c|}{$k$-higher} & \multicolumn{3}{c|}{$Ec$-lower} & \multicolumn{3}{c|}{$Ec$-higher} \\ \cline{2-16}
 & WS & ER & BA & WS & ER & BA & WS & ER & BA & WS & ER & BA & WS & ER & BA \\ \hline
\multicolumn{1}{|l|}{ WS } & 0.00 & 0.00 & 0.32 & 0.01 & 0.00 & 0.42 & 0.40 & 0.11 & 0.95 & 0.00 & 0.00 & 0.26 & 0.16 & 0.08 & 0.76 \\ \hline
\multicolumn{1}{|l|}{ ER } & 0.00 & 0.00 & 0.23 & 0.00 & 0.00 & 0.29 & 0.51 & 0.23 & 0.94 & 0.00 & 0.01 & 0.21 & 0.24 & 0.13 & 0.72 \\ \hline
\multicolumn{1}{|l|}{ BA } & 0.04 & 0.03 & 0.14 & 0.03 & 0.04 & 0.10 & 0.33 & 0.42 & 0.63 & 0.20 & 0.23 & 0.39 & 0.98 & 0.93 & 0.97 \\ \hline
\end{tabular}
\caption{The fraction of combinations in which the entire $N_P$ was mounted, by considering a buffer size of 4 puzzle components. Rows are for $N_{RE}$ and the columns represent $N_P$.}\label{tab:buffer4}
\end{table*}

We also considered the case were the buffer size follows the nodes degrees. Note that since the networks have an average degree of approximately 4, the average buffer size of the nodes is also 4. The results for this case are shown in Table~\ref{tab:degreeBuffer}. Compared with the case were all nodes have a buffer size of 4, the fraction of entirely mounted puzzles is higher for all combinations of $N_{RE}$ and $N_P$. Also, when this strategy is employed, almost all puzzles are mounted when $N_{RE}$ follows the BA model. 

\begin{table*}[!htpb]
\begin{tabular}{l|c|c|c|c|c|c|c|c|c|c|c|c|c|l|l|}
\cline{2-16}
 & \multicolumn{3}{c|}{Random} & \multicolumn{3}{c|}{$k$-lower} & \multicolumn{3}{c|}{$k$-higher} & \multicolumn{3}{c|}{$Ec$-lower} & \multicolumn{3}{c|}{$Ec$-higher} \\ \cline{2-16}
 & WS & ER & BA & WS & ER & BA & WS & ER & BA & WS & ER & BA & WS & ER & BA \\ \hline
\multicolumn{1}{|l|}{ WS } & 0.10 & 0.03 & 0.34 & 0.01 & 0.02 & 0.23 & 0.81 & 0.62 & 0.99 & 0.01 & 0.04 & 0.12 & 0.83 & 0.58 & 0.99 \\ \hline
\multicolumn{1}{|l|}{ ER } & 0.04 & 0.07 & 0.21 & 0.04 & 0.01 & 0.20 & 0.94 & 0.81 & 1.00 & 0.03 & 0.03 & 0.07 & 0.93 & 0.86 & 1.00 \\ \hline
\multicolumn{1}{|l|}{ BA } & 0.98 & 0.98 & 0.99 & 1.00 & 0.97 & 0.99 & 0.99 & 0.98 & 1.00 & 0.99 & 1.00 & 0.99 & 1.00 & 1.00 & 1.00 \\ \hline
\end{tabular}
\caption{The fraction of combinations in which the entire $N_P$ was mounted, by considering a buffer size equal to the degree of the nodes in $N_RE$. The rows indicate the $N_{RE}$ model used and the columns indicate the models used for $N_P$.}
\label{tab:degreeBuffer}
\end{table*}

\begin{figure*}[!htpb]
  \centering
    \subfigure[Random]{\includegraphics[width=0.3\textwidth]{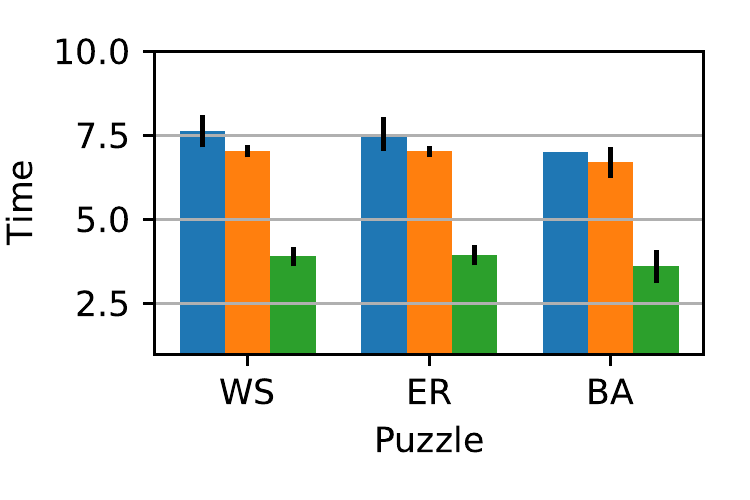}}
    \subfigure[$k$-lower]{\includegraphics[width=0.3\textwidth]{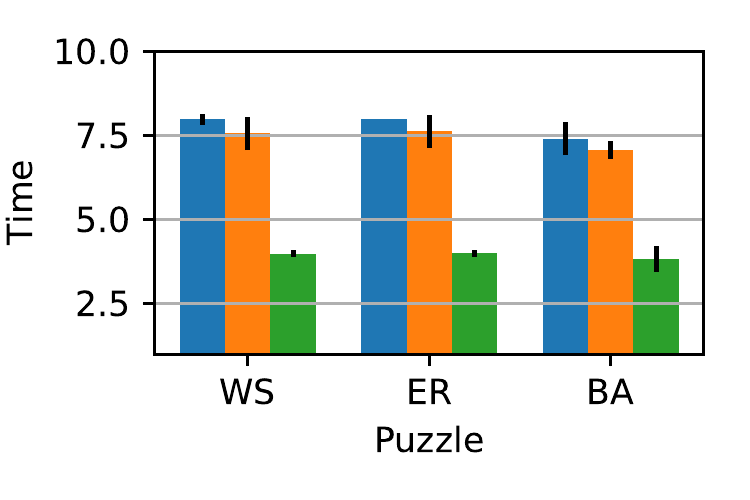}}
    \subfigure[$k$-higher]{\includegraphics[width=0.3\textwidth]{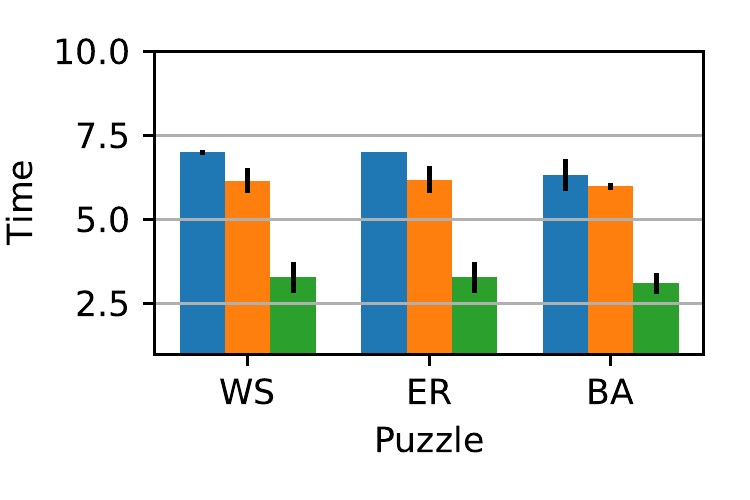}}
    \subfigure[$EC$-lower]{\includegraphics[width=0.3\textwidth]{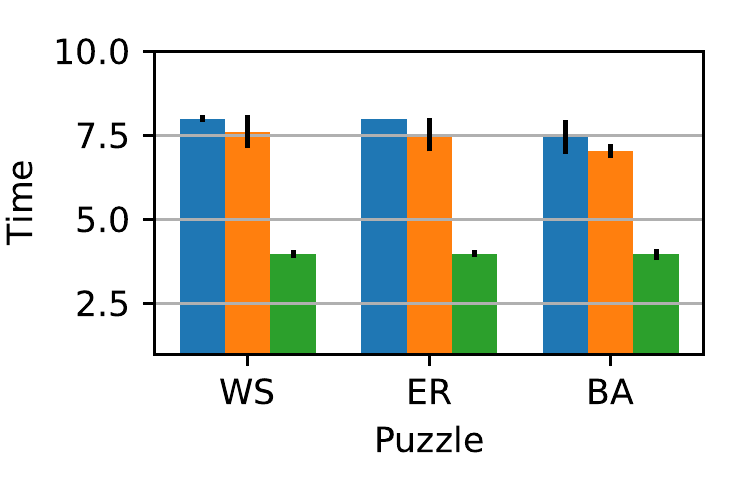}}
    \subfigure[$EC$-higher]{\includegraphics[width=0.3\textwidth]{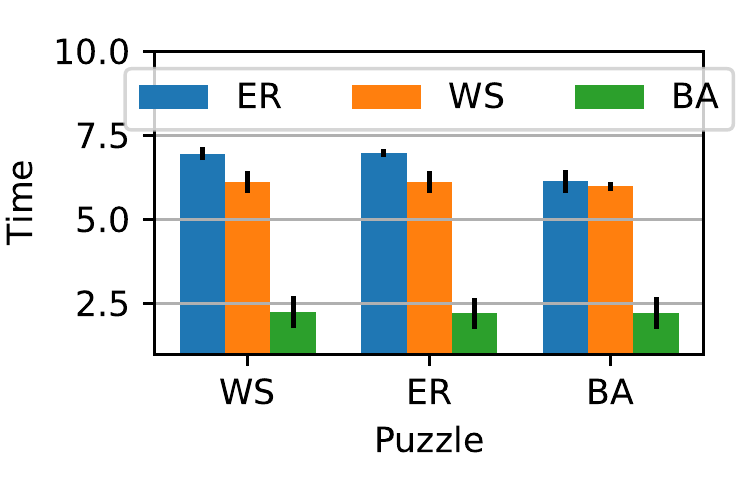}}
  \caption{Time to mount at least one $N_P$ with 90\% of the nodelets, by considering a buffer of size 4 for all nodes. Note that the colors represent the model used for the $N_{RE}$ networks.}
  \label{fig:mountedBar}
\end{figure*}

\begin{figure*}[!htpb]
  \centering
    \subfigure[Random]{\includegraphics[width=0.3\textwidth]{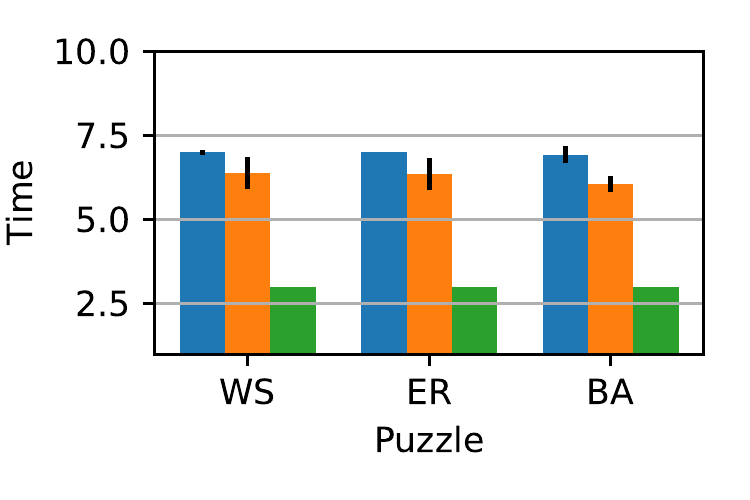}}
    \subfigure[$k$-lower]{\includegraphics[width=0.3\textwidth]{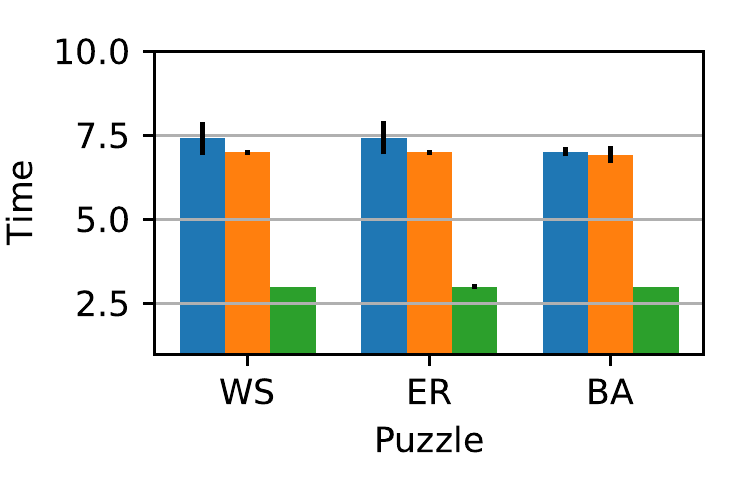}}
    \subfigure[$k$-higher]{\includegraphics[width=0.3\textwidth]{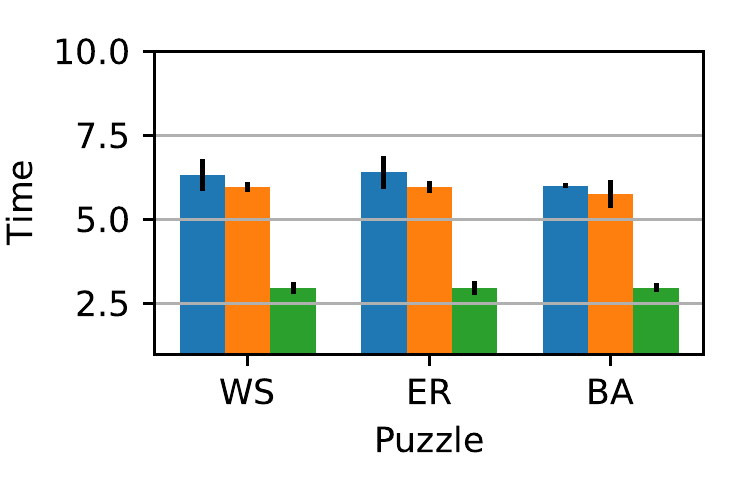}}
    \subfigure[$EC$-lower]{\includegraphics[width=0.3\textwidth]{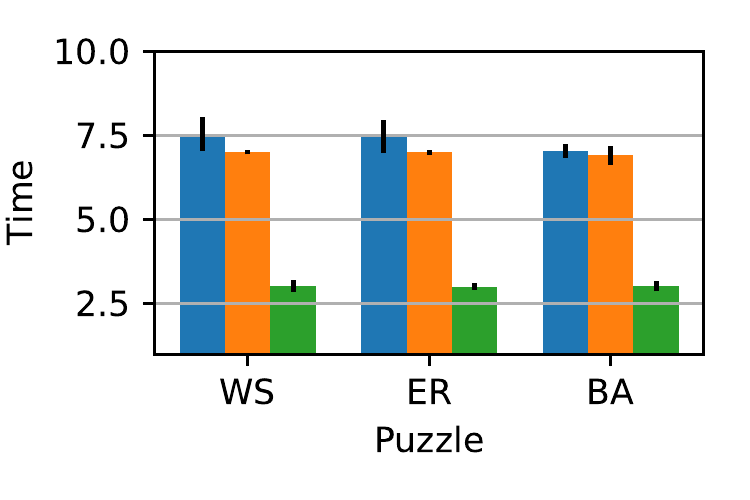}}
    \subfigure[$EC$-higher]{\includegraphics[width=0.3\textwidth]{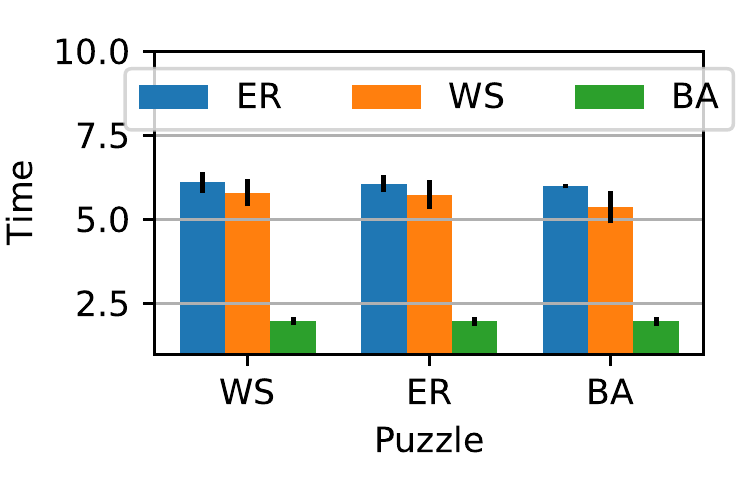}}
  \caption{Time to mount at least one $N_P$ with 90\% of the nodelets, by considering a buffer with size equal to the respective node degree. Note that the average buffer size is 4 for all the considered networks. Also, note that the colors represent the model used for the $N_{RE}$ networks.}
  \label{fig:mountedDegree}
\end{figure*}

\begin{figure*}[!htpb]
  \centering
    \subfigure[Random]{\includegraphics[width=0.3\textwidth]{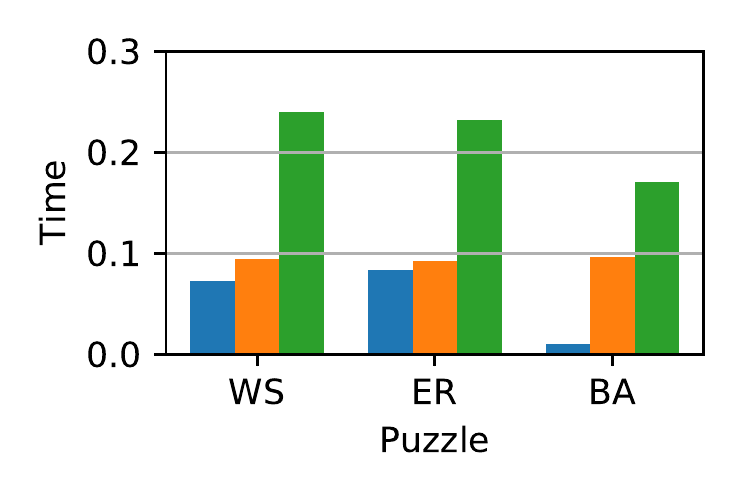}}
    \subfigure[$k$-lower]{\includegraphics[width=0.3\textwidth]{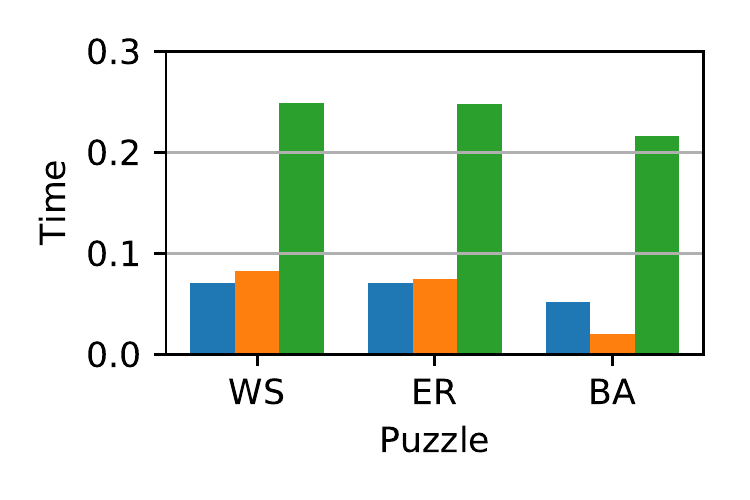}}
    \subfigure[$k$-higher]{\includegraphics[width=0.3\textwidth]{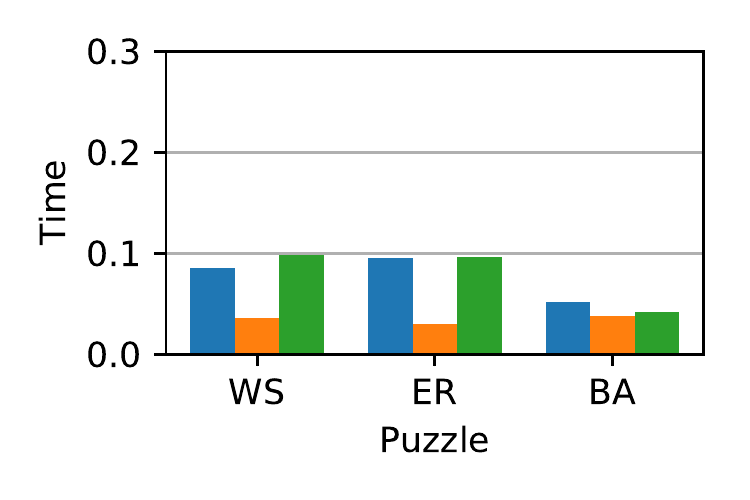}}
    \subfigure[$EC$-lower]{\includegraphics[width=0.3\textwidth]{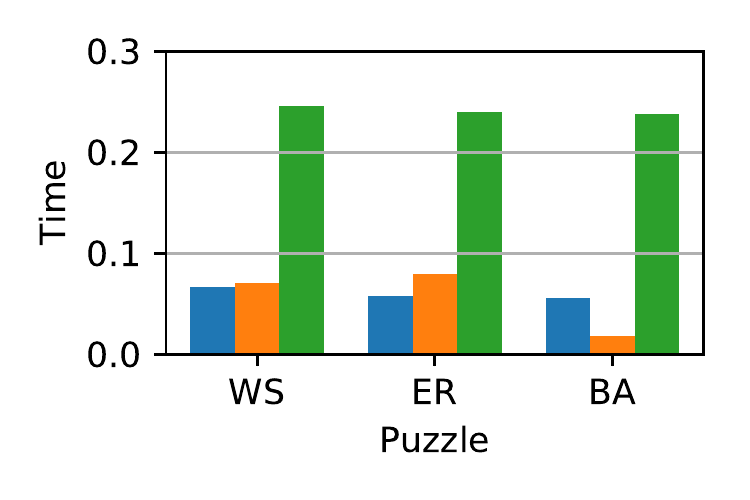}}
    \subfigure[$EC$-higher]{\includegraphics[width=0.3\textwidth]{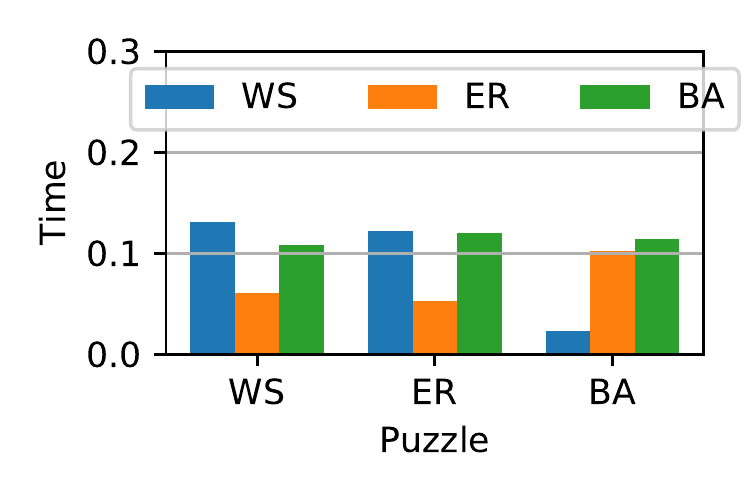}}
  \caption{Normalized difference between the time to mount at least one $N_P$ with 90\% of the nodelets. The difference is calculated by subtracting the times indicated in Figure~\ref{fig:mountedBar} by the times shown in Figure~\ref{fig:mountedDegree}, and dividing by times indicated in Figure~\ref{fig:mountedBar}. Note that the colors represent the model used for the $N_{RE}$ networks.}
  \label{fig:differences}
\end{figure*}

Another characteristic that can be measured is the time to mount the puzzle. Because in many cases the puzzle is not entirely mounted, we measured the time to mount 90\% of $N_P$ for both strategies: all nodes having the same buffer (Figure~\ref{fig:mountedBar}) and buffer size varying according to the $N_{RE}$ node degree (Figure~\ref{fig:mountedDegree}). For all tests, the second strategy leads to the puzzle being mounted faster. Furthermore, as observed in the case of infinite buffer size, the order of the most efficients $N_{RE}$ are: BA, ER, and WS. Figure~\ref{fig:differences} shows the normalized difference between the time to mount 90\% of $N_P$ for both strategies. Overall, larger differences are obtained when the BA network is employed as $N_{RE}$, except when the assigned priority is the highest values of $k$ and the $N_P$ is also a BA network. It is also interesting that the time difference also depends on $N_P$.

\section{Concluding Remarks}
Complex networks have been extensively studied mainly because of their inherent heterogeneity of connections.  Though research initially focused on the network topology, soon this problem was extended to include the relationship between topology and dynamics.  In the present work, we addressed the possibility to use complex networks as distributed systems processing artificial intelligence tasks, more specifically assembling puzzles that are complex networks themselves.  One of the immediate points of interest concerns the investigation of to which an extent the topology of the processing network influences the efficiency of achieving the solution, as expressed by the total assembling time.  

We analyze the proposed dynamics by considering memory with different sizes, called buffer. When an infinite size buffer was considered, the BA network was found to allow the shortest processing time, while the topology of the puzzle was not decisive on the assembling time. 
Another crucial characteristic that influenced the solution speed was the criterion adopted for the initial distribution of the puzzle pieces, here denominated assigned priority. All in all, improved performance was typically observed when the higher values of Eigenvector Centrality was used as the assigned priority.
A third interesting finding concerns the fact that assigning buffer sizes proportional to the degree of the respective nodes tended to improve the puzzle assembly performance as compared to using the buffer size equal to the average degree.

The reported work can be extended in several ways.  For instance, it could be assigned a probability for each node to discard pieces. Other possibilities include to test the effect of other topologies on puzzle assembling, and to consider other AI problems such as optimization and pattern recognition.

\section*{Acknowledgments}
Henrique F. de Arruda acknowledges the Coordenação de Aperfeiçoamento de Pessoal de Nível Superior - Brasil (CAPES) - Finance Code 001. Cesar H. Comin thanks FAPESP (grant no. 15/18942-8) for financial support. Luciano da F. Costa thanks CNPq (grant no. 307333/2013-2) and NAP-PRP-USP for sponsorship. This work has been supported also by FAPESP grants 11/50761-2 and 2015/22308-2. Research carried out using the computational resources of the Center for Mathematical Sciences Applied to Industry (CeMEAI) funded by FAPESP (grant 2013/07375-0).

\bibliography{references}
\end{document}